\documentstyle[12pt]{article}
\def\beq{\begin{equation}}
\def\eeq{\end{equation}}
\def\beqa{\begin{eqnarray}}
\def\eeqa{\end{eqnarray}}
\def\bc{\begin{center}}
\def\ec{\end{center}}

\def\bi{\begin{itemize}}
\def\ei{\end{itemize}}

\def\msol{M_\odot}

\begin{document}
{\normalsize
%\title{ MICROLENSING OF UNRESOLVED STARS AS A BROWN DWARF DETECTION METHOD

}

%\maketitle
{\bc{\bf {\large MICROLENSING OF UNRESOLVED STARS AS }}\ec}
%\vspace{0.1cm}
 {\bc{\bf {\large A BROWN~DWARF DETECTION METHOD}}\ec}
\vspace{1cm}
\bc{\sc {\large {Alain Bouquet, Jean Kaplan, Anne-Laure Melchior}}}\ec
\bc{\it{LPTHE, Universit\'es Paris 6 et Paris 7, unit\'e associ\'ee au CNRS
UA280, 75251 Paris Cedex~05, France}} \ec
\bc{\sc{\large{Yannick Giraud-H\'eraud} }}\ec
\bc{\it{Coll\`ege de France, laboratoire associ\'e au CNRS/IN2P3 (LA 41), 75231
Paris Cedex~05, France}}\ec
\bc{\sc{\large{Paul Baillon}}} \ec
\bc{\it{CERN, 1211 Geneva 23, Swizerland}}\ec

\vspace{0.5cm}
{\bc {Presented by Anne-Laure Melchior}\ec}
\vspace{0.5cm}
\bc {\sc ABSTRACT} \ec

{\footnotesize
\hspace{3pc}
{\parbox{30pc}{
We describe a project of brown dwarf detection in the dark halo of a galaxy
 using the microlensing effect. We argue that monitoring pixels instead of
stars could provide an enhancement in the number of detectable events.
We estimate the detection efficiency  with
a Monte-Carlo simulation. We expect a ten-fold increase with respect
to current experiments.
To assess the feasibility of this method we have determined the
photometric
precision of a pixel by comparing several pictures of a same field in the
 LMC.
} } }

%\section{Brown dwarfs as dark matter}
\vspace{1.2cm}
{\bf 1. Brown Dwarfs as Dark Matter}
%\subsection{\it Motivations}
\vspace{.5cm}

The dark halos of galaxies could be baryonic as suggested by the present
nucleosynthesis bounds{\small$^{1}$}\nocite{CARR}.
It may also explain
the disc-halo conspiracy observed in the rotation curves
{\small$^{2,3}$}\nocite{SANDERS,CASERTANO}.
Brown dwarfs, hypothetical compact hydrogen objects with a mass between
the evaporation limit{\small$^{4}$}{\nocite{DERUJULA}} $10^{-7} \,
\msol$  and the nuclear burning limit $10^{-1} \,
\msol$,  are a favoured possibility.

%Several experiments are engaged in the search of brown dwarfs.
%After briefly
%pointing out the different strategies of detection and their main features,
%we will focus on our proposal.
% \vspace{.5cm}

%\subsection{\it Direct detections}
%\vspace{1cm}
 Nearby brown dwarfs could in principle be directly detected by
 infrared obser\-vations{\small$^{5,6}$}\nocite{KERINS,DALY}.
Several groups are looking for individual brown dwarfs within our
galaxy.
 There
are already candidates in the Pleiades{\small$^{7}$}\nocite{SIMONS} which await
to be confirmed. This
procedure is not relevant for the dark matter in the halos, but it may
prove that brown dwarfs exist.
The halo of nearby galaxies might be detected by
satellites through the integrated infrared emission{\small$^{6}$}\nocite{DALY}.
These observations should be mostly sensitive to heavier brown dwarfs.

Brown dwarfs could also be indirectly detected by the  microlensing effect. A
star or a brown dwarf passing near the line of sight of a source,
deflects its light.
Although the angular deviation is too small to be
detected, the luminosity of the source is temporarily
increased{\small$^{8,9,10,11}$}\nocite{CHANG,GOTT,PACZYNSKI,GRIEST}.
 This amplification provides a
clear
signature~: the shape of the light curve should be achromatic, symmetric in
time and not repeated. The microlensing phenomenon has been
extensively
presented in this conference{\small$^{12,13}$}\nocite{JETZER,SPIRO}.

Microlensings may have already been observed on quasars, sitting behind a
gala\-xy{\small$^{14,15,16,17}$}\nocite{NOTTALE,IRWIN,CORRIGNAN,SCHILD}. Flux
variations
attributed to microlensing of the quasar by individual starlike objects of the
galaxy are expected to occur at a
rather high rate~: once about every few years. It is,
however, difficult to distinguish between ordinary stars and brown dwarfs
through this effect.

As suggested by Paczynski{\small $^{10}$}\nocite{PACZYNSKI}, microlensing
events may
be observed on stars of nearby galaxies. The rate of events
  per target is
much lower (typically once every million years), but millions of
 stars can be followed.
%\vspace{0.5cm}
Experiments are now running, using stars in the Large Magellanic Cloud
(LMC) as targets
(MACHO{\small $^{18}$}\nocite{MACHO}/EROS{\small $^{19}$}\nocite{EROS}).
These experiments are sensitive to  brown dwarfs
over the whole mass range. Notice that this
method is the only one to be sensitive to very low mass brown dwarfs.\\

%\vspace{0.5cm}
This paper proposes an alternative method to detect microlensing
events. Monitoring a galaxy by one million pixels instead of individual stars
could
significantly improve the detection efficiency.

\vspace{1cm}
%\section{Why monitor pixels rather than stars ?}
{\bf 2. Why Monitor Pixels rather than Stars ?}
\vspace{.5cm}

 The typical number of resolved stars is about 10$^5$ for
M31 and a few 10$^6$ in the LMC, which may at most give rise to a few
microlensing events per year. These galaxies, however, contain about $
10^{10-11} $
stars so that the potential number of sources is quite large.

To take advantage of all these potential targets, we
propose{\small $^{20}$}\nocite{BAILLON} to monitor
the light flux received by every pixel on the picture of a galaxy rather
than
the flux of the individual stars. In this way we will be able to detect the
flux
increase
 due the lensing of one
of the (many) stars present on the pixel even if it is not resolved.

Still, all lensings will not be detectable~: only those of the brightest
stars or those which produce a high amplification. High amplifications
mean a close approach of the lens to the line of sight and therefore seldom
occur.

With our proposed method, using the LMC as target we expect to gain a factor
around 10 on the number of events, compared to ongoing experiments.
This is illustrated on the following plot (figure 1) which displays the number
of events we expect to detect
with our method, as a function of the magnitude of the lensed stars.
These numbers have been obtained, in the conditions of the EROS experiment,
using a Monte~Carlo simulation presented in more details below.

 The limiting (visual) magnitude of the EROS experiment is around 19, and is
indicated by a vertical line.
The  area under the curve, marked ``stars'', corresponds to
microlensings of  monitored stars, whereas the area marked
``pixels'' corresponds to unresolved stars, the microlensings of which can
only be detected through our method. We can see clearly the proportion of the
resolved stars with
respect with the unresolved ones, and the appreciable gain in the
number of expected events.

\vspace{7cm}
{\bc Figure 1 : Magnitude of (unresolved) lensed stars \ec}
\vspace{1cm}

Another advantage of our method is that M31, where the number of resolved
stars is limited, becomes a promising target{\small
$^{21,20}$}\nocite{CROTTS,BAILLON}.
The expected number of events is similar to the LMC target. Individual stars in
M31 are fainter, so that higher and therefore rarer amplifications are needed
for a star to stand out of the background, but this is compensated by the
larger number of potential target stars in the field. Moreover, M31 possesses
its own dark
halo, the brown dwarfs of which also act as lenses.

\vspace{1cm}
%\section{How do we estimate the gain ?}
{\bf 3. How do we estimate the gain ?}
\vspace{.5cm}

% We study for a given experimental configuration if it could be possible
%to decrease
%the threshold of the detection or in others words to increase the number of
%stars to take into account.
% We plan to take pictures of the same field of a nearby galaxy with a
%minimum of pointing errors.
% Then we have to select events which stand above the fluctuations of the
%background with a given criterium.
%The next step will be to discriminate among these events the real
%microlensing ones.

We have estimated through a Monte-Carlo simulation the number of events we
expect to be able to detect with a monitoring of the pixel luminosity.
We present here our study under the conditions of the EROS experiment
{\small $^{19,13}$}\nocite{EROS,SPIRO} in order to test the feasibility of our
method.

We considered two galaxies of the
 Local Group : the L.M.C. (distance :
50kpc, angular extension : a few degrees, visible from the
South hemisphere) and M31 the largest galaxy of the Local Group (700kpc,
a few degrees, North Hemisphere). For both targets, we took
the luminosity function of the stars from the literature when
available, and used that of the solar neighbourhood otherwise (see Ref.
$^{\nocite{BAILLON}}$ for details).

Then, we  selected at random a brown dwarf in the halo of the Milky Way and of
the target galaxy, according to a simple halo model (density
decreasing as $1/(a^2+r^2)$, Maxwellian velocity distribution, identical halo
for M31 as for our galaxy, no halo for the LMC).
 We also selected at random a star in the target galaxy, with a weight
proportional to the product of the surface luminosity of the target area with
the star luminosity function.
We then computed the amplification of the lensed star as a function of
time.

If the amplification gets large enough, the light flux reaching the pixel in
the direction of the star will temporarily rise above the fluctuations of the
output of the pixel. If this rise is large enough, and lasts long enough  we
call this an event (see an example on figure 2).

\vspace{7cm}
{\bc Figure 2 : Aspect of a detected microlensing event \ec}
\vspace{1cm}

The fluctuation of the pixel output can prevent the detection of an event.
This fluctuation, defined as its standard deviation, includes several
contributions:

i) The statistical fluctuation of the background light : the night sky
luminosity plus the surface brightness of the target galaxy, corrected of
course for the absorption by the atmosphere, the mirrors and filters and for
the quantum efficiency of the CCD.

ii) The readout noise of the CCD.

iii) The residual errors due to imperfect matching of consecutive pictures.

We called an event a rise of the luminosity of a definite pixel  above 3
$\sigma$ during 3 consecutive exposures
 reaching 5 $\sigma$ in at least one of them. We expect that fluctuations can
only simulate such an ``event'' once every 50 years.

With the above requirements, for 120 nights 6 hours long, taking 15 minutes
exposures, and a seeing around 2", we expect the number of events displayed in
figure 3 as a function of the brown dwarf mass assumed here to be the same for
all brown dwarfs of the halo.
Of course we do not expect this to be true, but we know nothing about their
mass distribution. This presentation then allows to assess the sensitivity to
various brown dwarf masses.

%In our evaluation , all brown dwarfs were taken with the same mass.
\vspace{7cm}
{\bc Figure 3 : Expected number of microlensing events \ec}
\vspace{1cm}

The main limitations of this method come from the photometric precision
one can achieve, and from the variable stars which will be the main background.

\vspace{1cm}
%\section{Photometric precision}
{\bf 4. Photometric Precision}
\vspace{.5cm}

The number of detections expected depends crucially on the size of the
fluctuations of the pixel output. Since it can never be smaller than the
fluctuation of the number of photons reaching the pixel, the optimal situation
is obtained when all other sources of fluctuations are of the same order of
magnitude.
In our simulation statistical fluctuations were of the order of a few percent,
and we boldly assumed that all residual errors, the main source of which is the
matching in position and intensity of successive pictures, could be kept at the
same level. We have to check whether this is supported by real observations.

To this aim, the EROS collaboration provided us with a few pictures of the LMC
in two bands (non-standard B and R filters).
We studied 9 blue pictures of 100"$\times$100", with a seeing around 2" and a
pixel size of 1". We find that the relative fluctuations are smaller
than 4$\%$. The mean value of the pixel for these pictures is about 1700
photons
so that the corresponding statistical fluctuations are about 2.5 $\%$, and
therefore the residual errors are around 3\%.

To check the efficiency of the picture-matching algorithm
and to control the inputs of the Monte~Carlo  simulation, we elaborate some
synthetic pictures adjustable to different targets (M31/LMC), luminosity
function, experimental set-up, seeing, etc...
We generate some random fields of stars and evaluate the fluctuations.
This is a way to understand the different components of this fluctuation
(statistics, pixel matching, noises, etc...).
The results are compatible with the values measured on real
pictures (see figure 4).

\vspace{7cm}
{\bc Figure 4 : Distribution of the relative rms of the pixel response
estimated on 9 consecutive pictures (real and synthetic). \ec}
\vspace{1cm}

It is encouraging to note that, with observations not optimised for our
experiment, we find a rather good photometric accuracy. So that the number
of events we have evaluated seems valid. Of course, this evaluation must be
confirmed with a larger sample of pictures.

\vspace{1cm}
%\section{Variable stars}
{\bf 5. Variable Stars}
\vspace{.5cm}

Variable stars constitute the main source of background events.
As the selection criterium for the fluctuation requires an important signal,
this method can also distinguish the events which are symmetric in time and
unique. The achromaticity is more difficult to check in our approach because
the color of the lensed star may be different from the background color.
Nonetheless, many events in our simulation stand out both in blue and in red,
so one can study the time evolution of the event in both colors. In particular,
the ratio between the light flux {\it increases} of the pixel in the blue and
red  bands should be constant in time.

Note that our method is
sensitive to rather faint stars for a nearby target such as the LMC. Variable
stars are mostly
concentrated in a small region of the HR diagram~: they are bright red
stars. We expect a smaller relative number of variable stars in a larger
magnitude sample and thus less background than the running experiments.
This argument doesn't hold for more distant galaxies like M31 where detectable
events occur on brighter stars.

Nevertheless a target such as M31 seems promising in so far as
a special signature is expected{\small $^{ }$}\nocite{CROTTS}:
as it is a spiral galaxy tilted with respect to the line of sight, we  expect
more lensing events in the far side of the disk, which lies behind a larger
part of M31's halo.

\vspace{1cm}
%\section{What's next?}
{\bf 6. What's next~?}
\vspace{.5cm}

In a  first step we have to check on a larger sample of pictures that we can
reach the required photometric accuracy.

Then we could reanalyze the data of ongoing experiments$^{\nocite{EROS,MACHO}}$
following our approach. If we are right, the important gain in statistics will
allow to put constraints on the brown dwarf distribution in the halo.\\

Our simulation show that significant improvements in efficiency can be obtained
in several ways :

All our estimates have been made with a seeing of 2" where 17\% of the
starlight reaches the pixel at the center of the seeing spot.
 For a seeing of 1", 50\% of the starlight reaches this pixel, and our
sensitivity then
increases because the required amplifications  are smaller.
First estimates indeed indicate an enhancement of a factor 2 for the LMC and 5
for M31 with a seeing of 1".

To increase the sensitivity to larger brown dwarf masses one will have to
resort to multi-field procedures. Note that M31 seems a better target in this
respect (see figure 3). To support our conclusions we plan to make test
experiments on M31, using the Pic du Midi or CERGA telescopes.

On the other hand for small masses the shape of the amplification curve will be
difficult to observe as most events will last about 24 hours but will be
observed only during the night. This difficulty could be overcome by
correlating observations of telescopes in faraway sites.

On the basis of the above remarks, we have to define an optimal observation
device.

As an alternative to the CCD camera one can use of a photomultiplier array,
which has a far better photometric precision, but at the expense of a poorer
angular resolution{\small $^{20}$}\nocite{BAILLON}.

\vspace{1cm}

{\bf 7. Acknowledgements}
\vspace{0.5cm}

The authors wish to thank the EROS Collaboration for providing some
pictures of the LMC and in particular Reza Ansari, Christophe
 Magneville and Marc Moniez for their advices throughout this project.

%\bibliography{procrome}
%\bibliographystyle{unsrt}
%\begin{thebibliography}{}
%\bibitem{kn:CARR}
%\end{thebibliography}

\vspace{1cm}
{\bf 8. References}
\vspace{0.5cm}

%\bibitem{CARR}
1.~B.~J. Carr et~al,
{\em { Astrophysical Journal }}, {\bf 277}:99, 1984.

%\bibitem{SANDERS}
2.~R.H. Sanders,
{\em { Astronomy and Astrophysics Review}}, {\bf 2}:1--28, 1990.

%\bibitem{CASERTANO}
3.~S.~Casertano and J.H. Van~Gorkom,
 {\em { Astronomical Journal}}, {\bf 101}:1231, 1991.

%\bibitem{DERUJULA}
4.~A.~De~Rujula et al,
 {\em { Astronomy and Astrophysics}}, {\bf 254}:99, 1992.

%\bibitem{KERINS}
5.~E.J. Kerins and B.J. Carr,
{\it Monthly Not. R. Astron. Soc. },
{\it to be published}, 1993.

%\bibitem{DALY}
6.~R.A. Daly and G.C. Mclaughlin,
 {\em { Astrophysical Journal}}, {\bf 390}:423, 1992.

%\bibitem{SIMONS}
7.~D.A. Simons and E.E. Becklin,
{\em { Astrophysical Journal}}, {\bf 390}:431, 1992.

%\bibitem{CHANG}
8.~K.~Chang and S.~Refsdal,
{\em { Nature}},  {\bf 282}:561, 1979.

%\bibitem{GOTT}
9.~J.R. Gott~III,
 {\em { Astrophysical Journal}},  {\bf 243}:140, 1981.

%\bibitem{PACZYNSKI}
10.~B.~Paczy\'nski,
 {\em { Astrophysical Journal}}, {\bf 304}:1, 1986.

%\bibitem{GRIEST}
11.~K.~Griest,
 {\em { Astrophysical Journal}}, {\bf 366}:412--421, 1986.

%\bibitem{JETZER}
12.~P.~Jetzer,
 {\em  This conference}, 1993.

%\bibitem{SPIRO}
13.~M.~Spiro et~al,
 {\em This conference}, 1993.

%\bibitem{NOTTALE}
14.~L.~Nottale,
 {\em { Astronomy and Astrophysics}},  {\bf 157}:383, 1986.

%\bibitem{IRWIN}
15.~M.J. Irwin et~al,
 {\em { Astronomical Journal}}, {\bf 98}:1989, 1989.

%\bibitem{CORRIGNAN}
!6.~R.T. Corrignan et~al,
 {\em {Astronomical Journal}}, {\bf 102}:34, 1991.

%\bibitem{SCHILD}
17.~R.E. Schild,
 {\em { Astronomical Journal}}, {\bf 100}:1771, 1990.

%\bibitem{MACHO}
18.~D.P. Bennett et~al,
{\em  First data from the macho experiment},
to be published,~1993.

%\bibitem{EROS}
19.~E.~Aubourg et~al,
{\em The eros search for dark halo objects},
to be published, 1993.

%\bibitem{BAILLON}
20.~P.~P. Baillon et~al,
 {\em { Astronomy and Astrophysics}}, {\bf 277 (III)}:1, , 1993.

%\bibitem{CROTTS}
21.~A.P.S. Crotts,
 {\em { Astrophysical Journal}}, {\bf 399}:L43, 1992.

\end{document}